\documentclass{sf2a-conf2012}
\usepackage{graphicx}
\usepackage{hyperref}
\usepackage[]{natbib}  
\usepackage[cyr]{aeguill}
\usepackage{epstopdf}

\def\BibTeX{{\rm B\kern-.05em{\sc i\kern-.025em b}\kern-.08em
    T\kern-.1667em\lower.7ex\hbox{E}\kern-.125emX}}
\bibpunct{(}{)}{;}{a}{}{,}  

\begin{document}

\TitreGlobal{Reduction and Analysis of MUSE data}


\title{Reduction and Analysis of MUSE data}

\runningtitle{Reduction and Analysis of MUSE data}

\author{J. Richard}\address{CRAL, Universit\'e Lyon-1 and CNRS-UMR 5574, 9 avenue Charles Andr\'e, 69561 Saint-Genis Laval Cedex, France}

\author{R. Bacon$^1$}

\author{P. M. Weilbacher}\address{Leibniz-Institut f\"ur Astrophysik (AIP), An 
der Sternwarte 16, D-14482 Potsdam, Germany}

\author{O. Streicher$^2$}

\author{L. Wisotzki$^2$}

\author{E. C. Herenz$^2$}

\author{E. Slezak}\address{Observatoire de la C\^ote d'Azur, Bd de l'Observatoire, BP4229, FR-06304 Nice Cedex 4, France}

\author{M. Petremand}\address{LSIIT, UMR CNRS 7005, University of Strasbourg, B S. Brant, BP10413, 67412 Illkirch, France}

\author{A. Jalobeanu$^4$}

\author{C. Collet$^4$}

\author{M. Louys$^4$}

\author{the MUSE and DAHLIA teams}

%

\setcounter{page}{237}


\maketitle


\begin{abstract}

MUSE, the Multi Unit Spectroscopic Explorer, is a 2nd generation integral-field spectrograph under final assembly to see first light at the Very Large Telescope in 2013. By capturing $\sim90000$ optical spectra in a single exposure, MUSE represents a challenge for data reduction and analysis. We summarise here the 
main features of the Data Reduction System, as well as some of the tools under development by the MUSE 
consortium and the DAHLIA team to handle the large MUSE datacubes 
(about 4$\times10^8$ pixels) to recover the original astrophysical signal.

\end{abstract}



\section{Introduction}

The Multi-Unit Spectroscopic Explorer (MUSE, \citealt{MUSE}) is a second generation instrument to be commissionned in 2013 on the Very Large Telescope (VLT, unit telescope UT4). It is 
an integral-field spectrograph operating in the visible wavelength range with two main modes of operation: the Wide Field 
Mode (1x1 arcmin field-of view, 0.2''/pixel) and the Narrow Field Mode (7.5x7.5 arcsec, 0.025''/pixel), both with a spectral resolution of 1800-3600. 

The instrument is specifically designed to exploit the capabilities of the VLT Adaptive Optics Telescope Facility (AO), which can use four laser guide stars and a natural guide star to apply corrections via a deformable secondary mirror. 
The main scientific goal of MUSE, to be carried out during the Guaranteed Time Observations, will be to study the high redshift Universe through the measurement of the Lyman-$\alpha$ signature in distant galaxies  ($2.8<z<6.7$). Examples of other science topics include the study of nearby galaxies and intermediate redshift  groups (up to $z=0.8$) as well as globular clusters.

\begin{figure}[ht!]
 \centering
 \includegraphics[width=0.6\textwidth,clip]{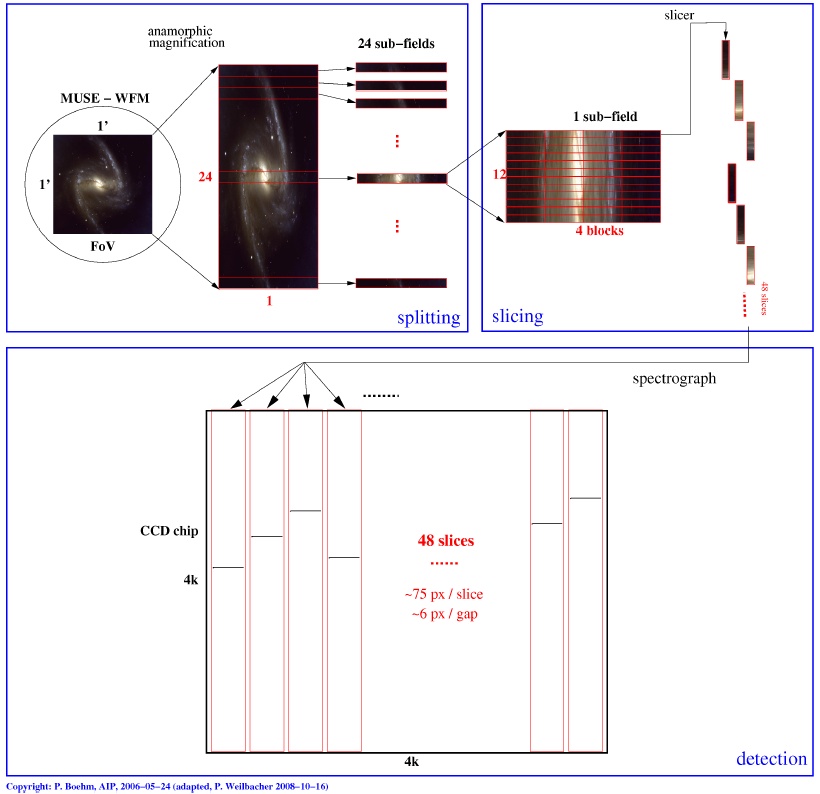}      
  \caption{(adapted from \citealt{DRS}) Splitting of MUSE data: example given for the case of the Wide-Field Mode (WFM). The field splitter separates the 1$\times$1 arcmin field-of-view into 24 sub-fields (top-left), each entering a separate IFU and sliced into 48 pseudo-slits (top-right). The various spectra produced are imaged onto a 4k x 4k detector (bottom).}
  \label{richard:fig1}
\end{figure}

In order to fully cover such a large field-of view, the instrument is composed of 24 integral-field units, each observing in parallel a 24$^\mathrm{th}$ of the field. The light entering each unit is sliced and reimaged into 48 pseudo-slits, then propagated through a spectrograph before reaching a 4k x 4k detector (Fig. \ref{richard:fig1}).

In total, the data acquired by MUSE in a single exposure amounts for 86400 optical spectra, or 
3.6$\times$10$^8$ pixels. A typical deep field observation with MUSE will have to combine 
$\sim$80 such exposures: this represents a challenge for data reduction and analysis. 
 
\section{MUSE data reduction}


A dedicated Data Reduction System (DRS) is being developed by the MUSE consortium 
under supervision of AIP, and will be delivered to ESO together with the MUSE 
instrument. Most of the functionalities described hereafter are now fully 
functional or at the stage of final tests / improvements \citep{DRS2}. 

The main task of the DRS is to convert the raw data coming from the 24 CCDs and create a fully 
calibrated datacube (in spatial, wavelength and flux coordinates), corrected for all sorts of 
instrumental and atmospheric effects. The system is designed to be embedded into the ESO data flow system and work as an automated pipeline interfaced 
with the ESO tools esorex and gasgano, as for other second-generation instruments (e.g. XShooter). It is written in the C language using the ESO Common Pipeline Library. 

In order to reach the best quality and reliability of the reduced data, only 
one interpolation step is performed at the end of the reduction to produce 
the final datacube. During the intermediate steps, the information from the 24 CCDs is propagated, for each pixel, in a master table called \textit{pixel table}.
Another important goal of the 
DRS is to propagate, throughout the data reduction steps, the error information corresponding to each pixel. 

\subsection{Data Reduction Cascade}

The first steps of the DRS perform a classical reduction of each of the 
24 CCDs: 

\begin{itemize}
\item{Creation of master bias, dark and flat-field frames.}
\item{Creation of master tables for geometrical, tracing and wavelength 
calibrations.}
\item{Then each science frame is corrected for bias, dark, flat 
and the \textit{pixel table} is created with the pixel values and 
the corresponding positions and wavelengths.}
\end{itemize}

In the second part of the data reduction cascade, the sky subtraction 
is performed on the entire pixel table, then the flux and astrometric 
calibrations are computed. Finally, the datacube (2 spatial axes and 
one wavelength axis) is constructed by a single interpolation of the 
pixel table. 

\subsection{Sky Subtraction}

Sky subtraction is a very important step in the reduction of datacubes, 
especially in the red part of the spectrum where a large number of atmospheric 
emission lines dominate the background noise. This is also the region where 
MUSE will observe faint and distant Lyman-$\alpha$ emitters. 

The current approach used in the DRS for sky subtraction is based on the 
modelling of the night sky (before resampling) into emission lines and 
continuum in single slices. This assumes a solid knowledge of the 
Line Spread Function (LSF) for each slice, which is modelled following the 
wavelength calibration using the brightest and most isolated lines of the 
arc exposures. More details on this sky 
subtraction method are presented in \citet{Streicher}.

An alternative sky subtraction procedure is currently being tested. The 
approach used is perpendicular to the previous one, in the sense that 
the sky fit and subtraction is performed in the spatial (rather than the 
spectral) direction. Bright continuum objects, emission lines and cosmics 
 are iteratively masked and removed from the fit. The main advantages of this 
classical sky subtraction is the low computing cost, and the fact that it does 
not depend on the knowledge of the LSF. However, it could be 
more vulnerable to the presence of bright objects in the case of crowded 
fields.


\section{MUSE data analysis}

\subsection{Fusion of MUSE data}

One of the main difficulties of the MUSE data analysis is to combine, in an optimal way, 
the signal obtained in multiple exposures taken under different atmospheric (seeing, 
transparency, sky variation) and instrumental (LSF,field location and orientation) conditions. 
In the DRS, this task is performed by directly interpolating multiple pixel tables into a 
single combined datacube during the final reduction step.

Alternatively, a more complex \textit{fusion} task (named HyperFusion) has been developed in the DAHLIA group \citep{Petremand}. A combined datacube is reconstructed by maximizing, in an optimal Bayesian context, a posterior probability built from the whole set of observations and their acquisition parameters.

\begin{figure}[ht!]
\begin{minipage}{8cm}
 \includegraphics[width=0.8\textwidth,clip]{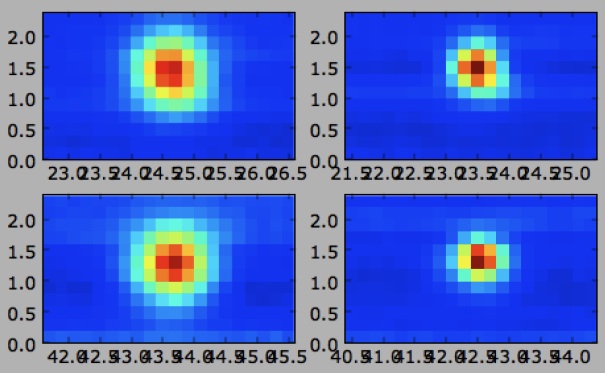}      
\end{minipage}
\begin{minipage}{8cm}
  \caption{Result of the fusion of MUSE datasets for 2 resolved sources using the direct method (left images) or the Bayesian 
fusion (right images).}
\end{minipage}
  \label{richard:fig2}
\end{figure}

Both the spatial (PSF) and spectral (LSF) 
responses are taken into account in this inverse problem. Compared to the ``direct'' approach, 
the \textit{fusion} clearly improves the 
resolution of astrophysical sources (Figure \ref{richard:fig2}), but this is a time-consuming process: combining 10 exposures take up to 5 days, compared to 2 hours with the DRS.

\subsection{Visualisation of MUSE data}

A dedicated tool, called \textit{QuickViz} \citep{Quickviz}\footnote{
publically available at
\url{http://lsiit-miv.u-strasbg.fr/paseo/cubevisualization.php}
}
 has been developed to help the MUSE 
users analyse such large datacubes. \textit{QuickViz} is designed as 
a plugin for Aladin and is written in the Java language.

 Specific features of 
\textit{QuickViz} are the following:

\begin{itemize}
\item{Coupled navigation between the spatial and spectral axes thanks both 
to calibrated cursors.}
\item{Full use of multi-core architectures to load and handle large datacubes 
as well as extract spectra from.}
\item{Data visualisation through multiple views, selections and simple processing algorithms.}
\item{Visualisation of the associated variance to a datacube in the form of an animation.}
\end{itemize}

\subsection{Source detection and extraction}
\begin{figure}[ht!]
\begin{minipage}{10cm}
 \includegraphics[width=1.0\textwidth]{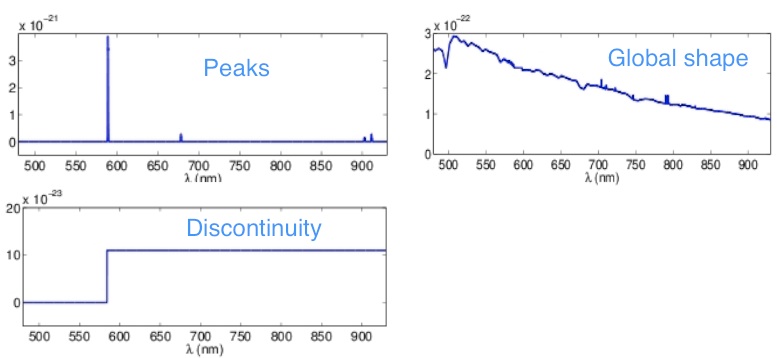}%
\end{minipage}
\begin{minipage}{7cm}
 \includegraphics[width=0.7\textwidth,clip]{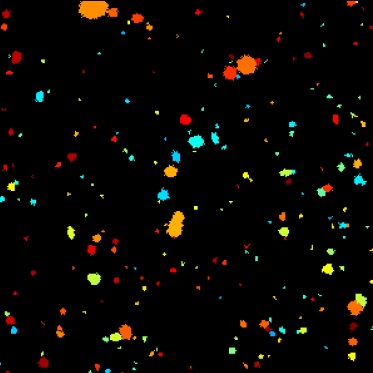}      
\end{minipage}
  \caption{{\bf Left:} Example of ``dictionary'' of spectral shapes used in the segmentation technique: a given source spectrum is fitted by the linear combination of such shapes. {\bf Right:} Example of datacube segmentation: sources are colored according to their spectrum.}
  \label{richard:fig3}
\end{figure}

The MUSE datacubes will contain a large number of sources (up to a few thousands in 
the deepest fields), appearing through their spectral continuum over a large wavelength range or through emission lines (only detected on a few wavelength slices). Several data analysis tools are currently being developed in order to produce a clean catalog of all detected objects in a given datacube:

\begin{enumerate}
\item{A classical approach uses an image analysis software (such as SExtractor) to produce a catalog of continuum sources from the \textit{white-light} image (obtained by collapsing the full datacube) as well as emission line sources 
(detected on narrow-band images over a few wavelength slices).}
\item{A similar approach performs the fitting and subtraction of all continuum sources before searching for emission lines by cross-correlation of the full datacube with a simple line model.}
\item{Alternatively, more complex data mining techniques are used to ``denoise'' the full datacube 
and identify sources based on a simple dictionary of spectral shapes (\citealt{Bourguignon}, Fig. \ref{richard:fig3}). These segmentation 
techniques can more easily de-blend overlapping sources.}
\item{Finally, a `perpendicular' technique is the use of point marked processes \citep{Chatelain} to identify 
sources assuming a simple morphological profile (elliptical shape, Sers\'\i c).}
\end{enumerate}

\begin{acknowledgements}
This work was partially funded by the French Research Agency (ANR) through the DAHLIA project (grant \#ANR-08-BLAN-0253).
\end{acknowledgements}

\bibliographystyle{aa}  
\bibliography{richard} 

\end{document}